# NOISE-BASED LOGIC GATES BY OPERATIONS ON THE REFERENCE SYSTEM

LASZLO B. KISH [1], WALTER C. DAUGHERITY [2]

[1] *Department of Electrical and Computer Engineering, Texas A&M University, College Station, TX 77843-3128, USA*

[2] *Department of Computer Science & Engineering, Texas A&M University, College Station, TX 77843-3112, USA*

We propose a new, low-complexity solution to realize multi-input-bit gates acting on exponentially large superpositions in noise-based logic processors. Two examples are shown, the NOT gate and the CNOT gate. The operations can be executed and repeated with polynomial time and hardware complexity. The lack of a solution of this problem had been one of the major issues prohibiting the efficient realization of Shor's algorithm by Instantaneous Noise-Based Logic, which runs on a classical Turing computer with a true random number generator. With the method described in this paper, we are one step closer to this goal.

*Keywords:* Quantum algorithms; noise-based logic; exponential speedup; low complexity; parallel operations.

## 1. Introduction

*1.1 On noise-based logic*

Noise-based logic (NBL) [1-14], including the brain [5-7], utilizes a reference system of uncorrelated (orthogonal) stochastic processes (noises), their superpositions, and their products to carry the logic information. There are many different realizations and types of NBL. Among them, the instantaneous NBL (INBL) schemes [3,4, 8-14] including the brain [5-7] do not require time averaging to identify the logic state and they produce their output superposition immediately. Practical examples are the random-telegraph-wave (RTW) based INBL schemes [3,4,7-14] described in subsections 1.2 and 1.3 below, that by utilizing *Achilles' heel logic functions* [2] are able to synthesize an exponentially large superposition of bit strings with polynomial hardware and time complexity, similarly to the original quantum computing schemes. Moreover, the single-bit quantum gate operations acting on exponentially large superpositions can also be implemented/run with polynomial complexity [9], indicating an exponential speedup due to the large parallelism represented by the exponentially large superposition. Thus there is hope that Shor's and other quantum algorithms can be run in an INBL-based processor, that is, in a classical binary computer with polynomial resolution of degree $2N$ and a physical random number generator [12]. Here $N$ is the required quantum bit resolution of the equivalent quantum computer. If such a goal is achieved, quantum algorithms could be realized in classical desktop and laptop computers, with the same computational complexity class as on a quantum computer, thus making today's computation-based encryption schemes insecure.

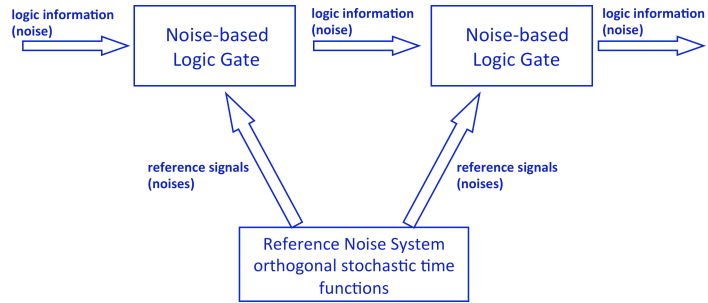

Fig. 1. Essential elements in a noise-based logic processor. The reference system supplies the $2N$ statistically independent noises (stochastic AC signals). The gates form the superpositions representing the logic states and act upon them.

However, the 2-bit gate operations, such as the CNOT, had been problematic on superpositions. In the most successful case to date [9], exponential speedup with polynomial complexity was attained by a single CNOT operation on an exponentially large, $O(2^N)$, superposition. However, each additional CNOT operation doubled the required hardware complexity, so cascaded CNOT operations led to exponentially large and thus infeasible hardware [9].

In the present paper, we introduce a new philosophy for the INBL gate operations on superpositions by implementing active circuitry manipulating the reference system that supplies the INBL processor with the reference noises. For the sake of simplicity of exposition, but without restricting generality, we present the simplest RTW-based INBL, even though many of its superpositions produce zero amplitudes most of the time. This characteristic makes the simplest scheme impractical and only academically/educationally interesting, as a toy model. However, it is straightforward to implement the new logic gates introduced in this paper using the more advanced/practical types of RTW-based schemes that avoid the zero-problem, namely, asymmetric [8], time-shifted [8,13] and complex RTWs [9], because the utilized features of RTWs are similar in each of these systems.

For the convenience of the Reader, in subsection 1.2 we briefly outline the noise-based product strings used to represent integer numbers and their superpositions in INBL. Then in subsection 1.3 we briefly describe the simplest scheme where the noises belong to the simplest type of RTWs. That system will be used to illustrate the new type of logic gate operations on superpositions that will be introduced in Section 2.

*1.2 The superposition of integer numbers in Instantaneous Noise-based Logic*

In a system with $N$ noise-bits [1], the reference base consists of $2N$ independent, uncorrelated noise sources, $W_{i,j}(t)$, each representing the fingerprint of a bit value $j \in \{0,1\}$, in the $N$ noise-bit system, $i \in \{0,...,N-1\}$. This system, through the superpositions of its bit strings (see Equation 1 below), can represent $2^{2^N}$ different logic values [1,2], that is, its logic depth corresponds to $2^N$ classical bits. Any integer number $R$ in the range of $\left[0, 2^{N-1}\right]$ is represented by its corresponding bit string, which is a new RTW, $X_R(t)$, formed by the product of the relevant $W_{i,j}(t)$ functions [2,7,11,13]:

*Noise-based parallel operations on the reference system for NOT and CNOT*

$$X_R(t) = \prod_{i=0}^{2N-1} W_{i,j(i,R)}(t) \ , \tag{1}$$

where the $j(i,R)$ function provides the value (0 or 1) of the *i*-th bit in the product string. Note that, for continuum noises, the bandwidth must be limited by filters or the natural bandwidth of the system, to avoid increasing the time complexity. This scheme, similarly to quantum computing schemes, is able to represent the *superposition* of states (signals) representing integer numbers in the form of a sum *Y*(*t*) of amplitudes of the relevant product strings:

$$Y(t) = X_{R1} + X_{R2} + ... \ . \tag{2}$$

A powerful feature of INBL that is the superposition of all integer numbers, that is, the universe $U(t)$, and many of its sub-superpositions, can be generated by utilizing Achilles' heel functions with polynomial complexity [2], (for concrete examples, see [13]), even though the sum representing the universe consists of an exponentially large ($2^N$) number of elements. For example, the universe is

$$U(t) = \prod_{i=0}^{2N-1} \left[ W_{i,0}(t) + W_{i,1}(t) \right] \ . \tag{3}$$

*1.3 Random telegraph waves as noise base*

In the rest of this paper, we deal with the simplest random telegraph waves. For more advanced RTW types [8,9,13], proper modifications of the new gates will be necessary. The situation also depends on the given application and may need additional steps.

The simplest RTW-based INBL system has a deterministic and identical clock for all the logic values. At the beginning of the clock period, the value of the RTW, -1 or +1, is randomly chosen with identical (0.5) probability and it remains static during the whole clock period. In this way, the time and amplitude complexity does not increase during multiplication. The essential properties are zero mean and orthogonality:

$$\langle W_{i,j}(t) \rangle = 0 , \qquad (4)$$

$$\langle W_{i,j}(t) W_{m,n}(t) \rangle = \delta(i,m)\delta(j,n) , \qquad (5)$$

where $\delta(r,s)$ is the Kronecker symbol. The product of two orthogonal RTWs is a new RTW, $[W_{i,j}(t) W_{m,n}(t)] = W_{p,q}(t)$, that is orthogonal to the original RTWs. With $i \neq m$ and/or $j \neq n$ this is shown by

$$\langle W_{p,q}(t) W_{i,j}(t) \rangle = \langle W_{m,n}(t) \rangle = 0 \quad \text{and} \quad \langle W_{p,q}(t) W_{m,n}(t) \rangle = \langle W_{i,j}(t) \rangle = 0 , \qquad (6)$$

so $W_{p,q}(t)$ is a new RTW that is uncorrelated with $W_{i,j}(t)$ and $W_{m,n}(t)$.

The NOT gate is a simple product operation

$$\text{NOT}_i = W_{i,j}(t) W_{i,k}(t)^* , \qquad (7)$$

where $\text{NOT}_i [W_{i,j}(t)] = W_{i,k}(t)$ with $j \neq k$, $j \in \{0,1\}$.

The NOT gate also works on any selected bit in a product strings or in a superposition, for example,

$$\text{NOT}_i [W_{1,a}(t) W_{2,b}(t) ... W_{i,j}(t) ... W_{N,z}(t)] = W_{1,a}(t) W_{2,b}(t) \, ... \, W_{i,k}(t) \, ... \, W_{N,z}(t) , \qquad (8)$$

*Noise-based parallel operations on the reference system for NOT and CNOT*

where $j \neq k$, $j \in \{0,1\}$.

We will show in Section 2 that the CNOT operations acting on arbitrary bits in a superposition can be built up by a combination of NOT gates acting on proper reference wires in a noise-based logic processor.

*1.4 Earlier attempts for achieving CNOT operations*

Many attempts were carried out, with only the version based on "ghost bits" being published [9], yet they all had a common problem. A single CNOT operation yielded the expected exponential speedup of $O(2^N)$ compared to classical binary logic gates because the gate simultaneously acted on all the product strings in the superposition. However, when the CNOT operation was repeated on various bit pairs, the hardware and/or time complexity grew exponentially with the number of the CNOT operations. Thus, after the CNOT gate was applied to each subsequent bit pair in the product string the complexity approached that of the classical computers.
The new scheme described in this paper avoids this problem.

## 2. The new NOT and CNOT gates

Figure 2 shows the generic noise-based logic processor scheme where the "Hyperspace Synthesizer/Processor" unit generates the given logic superposition state *Y* on which the subsequent logic gates will act. Examples are Equation 3 or its truncated versions to generate the universe or its fractions.

In such systems, we can do operations not only directly on *Y* but also on the reference wires. We show two examples: the realization of the NOT and the CNOT gates.

The new NOT gate is simple; it acts on the reference wires of the chosen bit by applying Equation 7 on both wires (which represent the 1 and 0 bit values). See Figure 3, in which the realization of the $NOT_3$ gate acts on the 3rd bit of each product string in the superposition *Y* by inverting the value of its 3rd bit. The operation represents a $O(2^N)$ speedup compared to classical computers because it acts on $O(2^N)$ number of orthogonal strings.

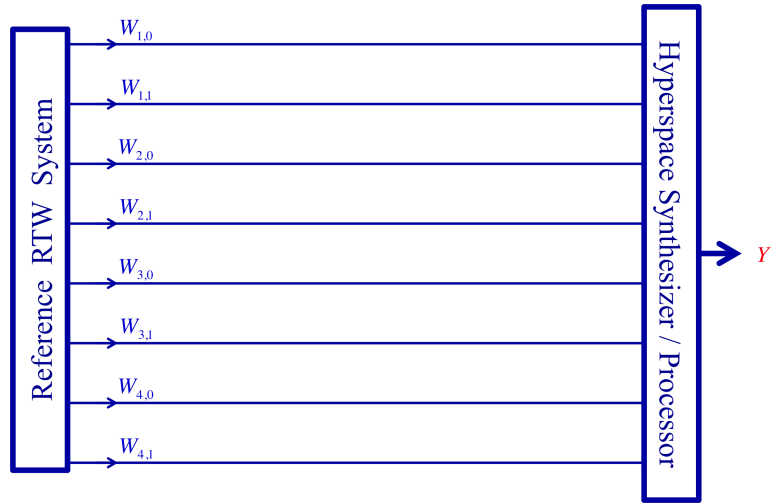

Fig. 2. The generic noise-based logic processor scheme generating the given logic superposition state *Y* on which the logic gates will act.

The realization of the CNOT gates requires similar techniques but is more complex when multiple gates are involved because the operations may be non-commutative. The CNOT gate inverts the target bit if the control bit has high value. The method is as follows:

a) First, the NOT operator of the target bit is inserted into the reference wire carrying the value 1 of the control bit. This circuitry will invert the target bit in all the product strings that contain value 1 of the control bit.

b) Then, whenever it is necessary (such as with a sequence of CNOT operations which is not commutative), add the proper NOT operators applied to other reference wires to correct for excess NOTs and to supply missing NOTs. The examples below demonstrate this straightforward procedure.

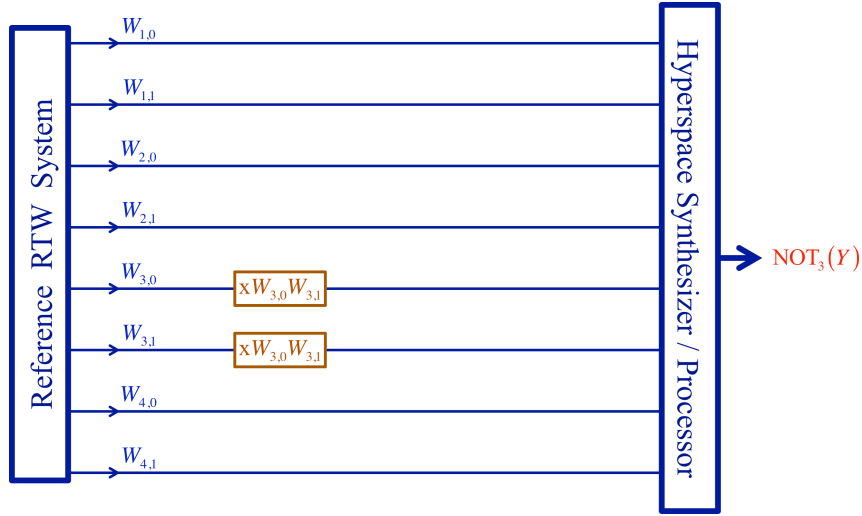

Fig. 3. Implementation of the $\text{NOT}_3(Y)$ gate; see Equation 7. It acts simultaneously and instantaneously on each product string in the $\text{O}(2^N)$ superposition that contains any value of bit 3 by inverting that value.

*2.1 Single CNOT operation*

Implementation of the $\text{CNOT}_{2,3}(Y)$ gate is displayed in Figure 4. It acts simultaneously and instantaneously on each product string in the superposition that contains value 1 of the control bit (bit 2) by inverting the value of the target bit (bit 3).

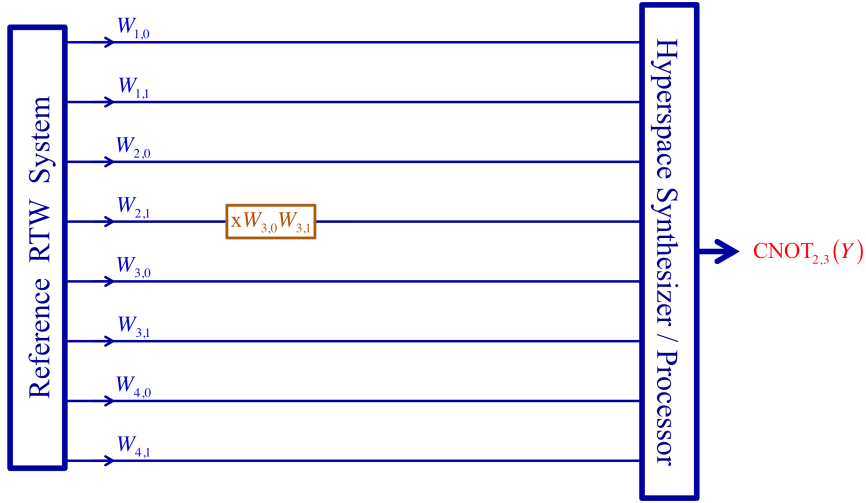

Fig. 4. Implementation of the $\text{CNOT}_{2,3}(Y)$ gate. It acts simultaneously and instantaneously on each product string in the superposition that contains value 1 of the control bit (bit 2) by inverting the value of the target bit (bit 3).

*2.2 Two or more cascaded CNOT operations*

In this case, we must consider whether or not the sequence of CNOT operations is independent. When the order of CNOT operations is such that they do not act on the control bit of the next CNOT, the NBL gate is simply the combination of the single CNOT operations shown above; see Figure 5, which shows the example of the $\text{CNOT}_{2,3}\text{CNOT}_{1,2}(Y)$ cascaded gates.

However, when the order of CNOT operations is such that a CNOT gate in the cascade acts on the control bit of a subsequent CNOT, that is, the gates "interact," then extra NOT operations must be added to achieve the correct result. See Figure 6, which shows the example of the $\text{CNOT}_{1,2}\text{CNOT}_{2,3}(Y)$ gate. We start with the circuitry of the non-interacting $\text{CNOT}_{2,3}\text{CNOT}_{1,2}(Y)$ gate (Figure 5). That will not yield the correct result for the target bit (bit 3) if the value of the first control bit (bit 1) is 1, so this problem is corrected by the $\text{NOT}_3$ operation inserted in reference wire $W_{1,1}$; see Figure 6.

*Noise-based parallel operations on the reference system for NOT and CNOT*

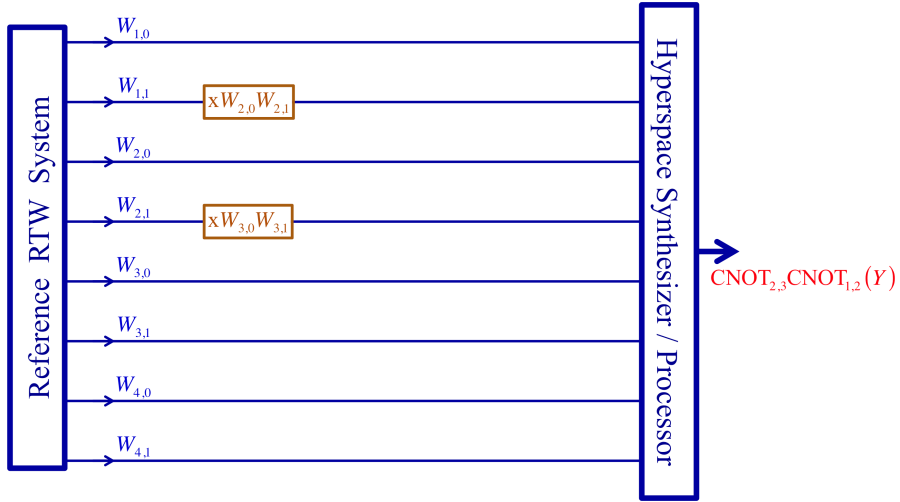

Fig. 5. The non-interacting cascade $\text{CNOT}_{2,3}\text{CNOT}_{1,2}$ gate.

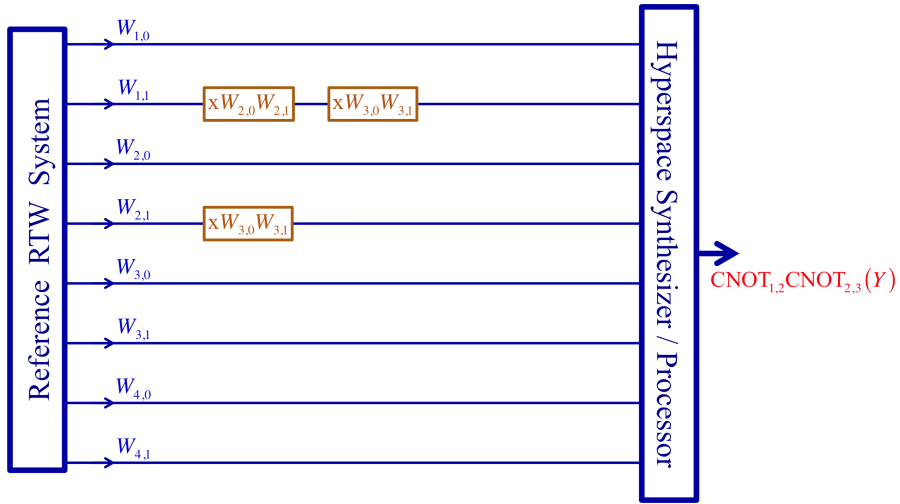

Fig. 6. The interacting cascade $\text{CNOT}_{1,2}\text{CNOT}_{2,3}$ gate.

Finally, Figure 7 presents the $\text{CNOT}_{3,4}\text{CNOT}_{2,3}\text{CNOT}_{1,2}(Y)$ as an example for another non-interacting CNOT cascade and, in Figure 8, the $\text{CNOT}_{1,2}\text{CNOT}_{2,3}\text{CNOT}_{3,4}(Y)$ is shown as another example for non-commutative interacting gates.

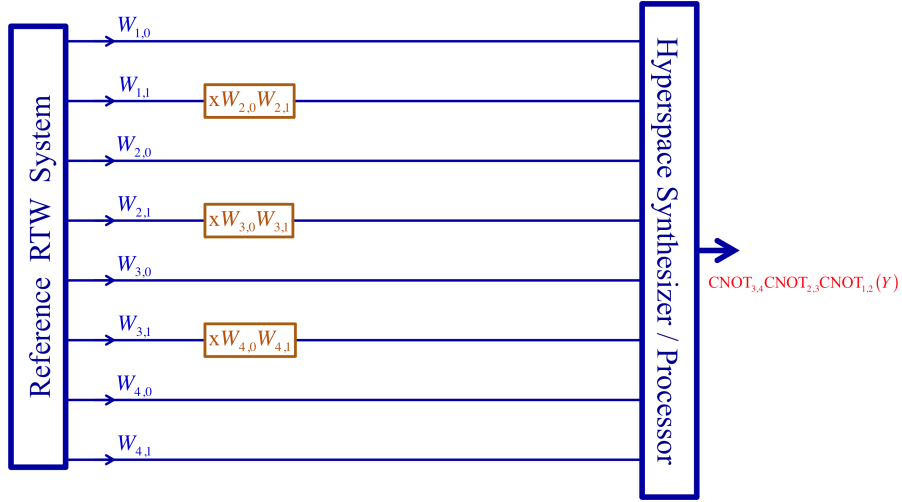

Fig. 7. The non-interacting cascade $\text{CNOT}_{3,4}\text{CNOT}_{2,3}\text{CNOT}_{1,2}$ gate.

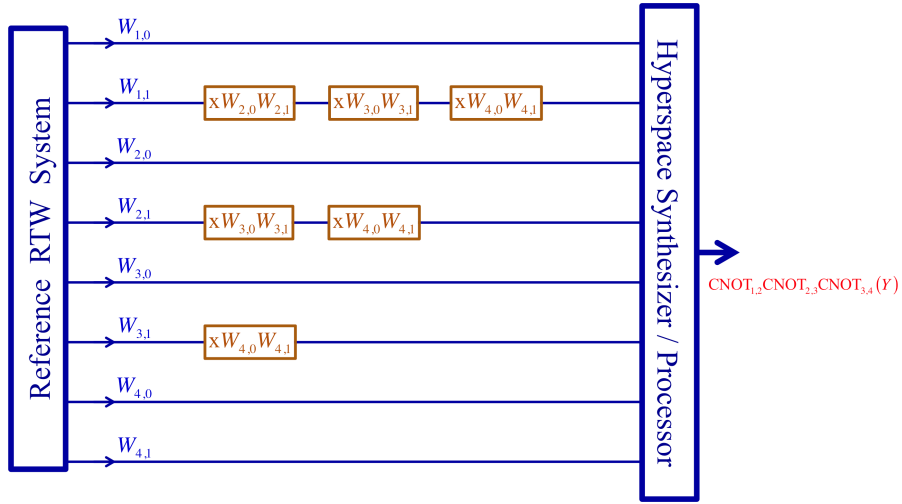

Fig. 8. The interacting cascade $\text{CNOT}_{1,2}\text{CNOT}_{2,3}\text{CNOT}_{3,4}$ gate.

*2.3 Hardware and time complexity*

These examples indicate that cascaded CNOT systems have only polynomial hardware complexity versus *N* while they are acting on exponentially large, $\mathrm{O}(2^N)$, superpositions of strings (that is, integer numbers) with time complexity $\mathrm{O}(1)$ because the result is instantaneously produced. These facts indicate exponential speedup of these operations compared to the case of classical digital computers. Our conjecture is that the number *M* of NOT hardware elements required to construct an *N* long cascade of CNOT gates is bounded as:

$$N \leq M \leq 0.5N(N+1) \ . \tag{9}$$

Note that though our conjecture is plausible and satisfied by the simple examples we have shown, a rigorous mathematical proof is missing at the moment.

**Conclusions**

A successful effort has been carried out to keep polynomial hardware and time complexity with cascaded CNOT parallel operations on exponentially large superpositions in Instantaneous Noise-Based Logic operating with classical physical elements and physical noise, that is, true random numbers. The lack of a solution of this problem had been one of the major issues prohibiting the efficient realization of Shor's algorithm by Instantaneous Noise-Based Logic, which runs on a classical Turing computer with a true random number generator. With the method described in this paper, we are one step closer to this goal.

We also note that the results may help to understand the wiring and signal structure of the brain [5-7] and its neural code.

**Acknowledgement**

Valuable discussions with Andreas Klappenecker are appreciated.

**References**

[1] L.B. Kish, Noise-based logic: binary, multi-valued, or fuzzy, with optional superposition of logic states, *Physics Letters A* **373** (2009) 911–918. http://arxiv.org/abs/0808.3162

[2] L.B. Kish, S. Khatri, S. Sethuraman, Noise-based logic hyperspace with the superposition of $2^N$ states in a single wire, *Physics Letters A* **373** (2009) 1928–1934. http://arxiv.org/abs/0901.3947

[3] L.B. Kish, S. Khatri, F. Peper, Instantaneous noise-based logic, *Fluctuation and Noise Letters* **9** (2010) 323–330. http://arxiv.org/abs/1004.2652

[4] F. Peper, L.B. Kish, Instantaneous, non-squeezed, noise-based logic, *Fluctuation and Noise Letters* **10** (2011) 231–237. Open access.

[5] F. Grueneis, "An interpretation of 1/f Fluctuations in Neuronal Spike trains," *Biol. Cybern.* **60** (1989) 161-169.

[6] S.M. Bezrukov, L.B. Kish, "Deterministic multivalued logic scheme for information processing and routing in the brain," *Physics Letters A* **373** (2009) 2338-2342.

[7] L.B. Kish, C.G. Granqvist, S.M. Bezrukov, T. Horvath, "Brain: Biological noise-based logic," The 4th International Conference on Cognitive Neurodynamics, June 2013, Sigtuna, Sweden. Proceedings: Advances in Cognitive Neurodynamics (IV), H. Liljenström (ed.) Springer, (2014) pp. 319-322. DOI: 10.1007/978-94-017-9548-7_45
http://link.springer.com/chapter/10.1007%2F978-94-017-9548-7_45

[8] H. Wen, L.B. Kish, Noise-based logic: Why noise? A comparative study of the necessity of randomness out of orthogonality, *Fluctuation and Noise Letters* **11** (2012) 1250021/1–1250021/7. http://arxiv.org/abs/1204.2545

[9] H. Wen, L.B. Kish, A. Klappenecker, F. Peper, New noise-based logic representations to avoid some problems with time complexity, *Fluctuation and Noise Letters* **11** (2012) 1250003/1–1250003/8. http://arxiv.org/abs/1111.3859

[10]  H. Wen, L.B. Kish, A. Klappenecker, Complex noise-bits and large-scale instantaneous parallel operations with low complexity, *Fluctuation and Noise Letters* **12** (2013) 1350002. http://vixra.org/abs/1208.0226

[11]  L.L Stachó, Fast measurement of hyperspace vectors in noise-based logic, *Fluctuation and Noise Letters* **11** (2012) 1250001.

[12]  L.B. Kish, S. Khatri, T. Horvath, Computation using noise-based logic: efficient string verification over a slow communication channel, *European Journal of Physics B* **79** (2011) 85–90. http://arxiv.org/abs/1005.1560

[13]  L.B. Kish, C.G. Granqvist, T. Horvath, A. Klappenecker, H. Wen, S.M. Bezrukov, "Bird's-eye view on noise-based logic," *International Journal of Modern Physics: Conference Series* **33** (2014) 1460363. Open access: http://www.worldscientific.com/doi/pdfplus/10.1142/S2010194514603639

[14]  B. Zhang, L.B. Kish, C.G. Granqvist, "Drawing from hats by noise-based logic," *International Journal of Parallel, Emergent and Distributed Systems* **32** (2017), 244-251; http://dx.doi.org/10.1080/17445760.2016.1140168 ; http://arxiv.org/abs/1511.03552.